\documentclass[letterpaper,twocolumn,ruled,vlined,linesnumbered,twocolumn]{IEEEtran}
\usepackage[utf8]{inputenc}
\usepackage{array}
\usepackage{verbatim}
\usepackage{multirow}
\usepackage{amsmath}
\usepackage{amssymb}
\usepackage{graphicx}
\usepackage{setspace}
\makeatletter

\pdfpageheight\paperheight
\pdfpagewidth\paperwidth

\providecommand{\tabularnewline}{\\}

\usepackage{amsmath}
\usepackage{amsfonts}
\usepackage{amssymb}
\usepackage{amsthm}

\usepackage{graphicx}

\usepackage{multirow}
\usepackage{array}
\usepackage{slashbox}
\usepackage{makecell}
\usepackage{threeparttable}

\usepackage{rotating}

\usepackage{pict2e}
\usepackage{cases}
\usepackage{enumitem}
\usepackage{footnote}
\makesavenoteenv{tabular}
\makesavenoteenv{table}

\usepackage{floatrow}
\usepackage[vlined,linesnumberedhidden]{algorithm2e}

\usepackage{color}
\usepackage{cite} 

\usepackage{url}
\newcommand{\figref}{Fig. }
\newcommand{\tabref}{Table }
\newcommand{\secref}{Section }

\title{\vspace{-6pt}A Two-staged Adaptive Successive Cancellation List Decoding for Polar Codes\vspace{-8pt}}

\author{ ChenYang Xia$^{1}$, YouZhe Fan$^{2}$, Chi-Ying Tsui$^{3}$\\ $^{1,3}$Department of ECE, the HKUST, HKSAR, China $^{2}$MaxLinear, Carlsbad, CA, USA\\$^{1,2}$\{cxia, jasonfan\}@connect.ust.hk, $^{3}$eetsui@ust.hk \vspace{-24pt}}

\@ifundefined{showcaptionsetup}{}{%
 \PassOptionsToPackage{caption=false}{subfig}}
\usepackage{subfig}
\makeatother

\begin{document}
\maketitle

\begin{abstract}
Polar codes achieve outstanding error correction performance when
using successive cancellation list (SCL) decoding with cyclic redundancy
check. A larger list size brings better decoding performance and is
essential for practical applications such as 5G communication networks.
However, the decoding speed of SCL decreases with increased list size.
Adaptive SCL (A-SCL) decoding can greatly enhance the decoding speed,
but the decoding latency for each codeword is different so A-SCL is
not a good choice for hardware-based applications. In this paper,
a hardware-friendly two-staged adaptive SCL (TA-SCL) decoding algorithm
is proposed such that a constant input data rate is supported even
if the list size for each codeword is different. A mathematical model
based on Markov chain is derived to explore the bounds of its decoding
performance. Simulation results show that the throughput of TA-SCL
is tripled for good channel conditions with negligible performance
degradation and hardware overhead. \vspace{-4pt}
\end{abstract}

\begin{IEEEkeywords}
Polar codes, Successive cancellation list decoding, Adaptive decoding,
Markov chain, Hardware-friendly algorithm\vspace{-6pt}
\end{IEEEkeywords}

\section{Introduction}

To improve error correction performance of polar codes \cite{earikan_bilkent_tit_2009_polar},
successive cancellation list (SCL) decoding \cite{ital_ucsd_tit_2015_list,kchen_bupt_iet_2012_lscd}
is the most popular decoding choice. $\mathcal{L}$ (called the list
size) successive cancellation (SC) decodings \cite{cleroux_mcgill_tsp_2013_semiparallel,yzfan_hkust_tsp_2014_effps}
are executed concurrently to decode a polar codeword and $\mathcal{L}$
candidates of decoded vectors are kept during decoding \cite{ital_ucsd_tit_2015_list,kchen_bupt_iet_2012_lscd}.
Compared with SC decoding, SCL decoding improves the error correction
performance as the probability of one of the $\mathcal{L}$ candidates
to be the correct decoded vector is higher, and a larger list size
brings a better error correction performance. %
In \cite{kniu_bupt_cl_2012_crc}, cyclic redundancy check (CRC) codes
are concatenated as outer codes with polar codes, and CRC is applied
to all the candidates to see whether any candidate is the valid decoding
output. From the experimental results presented in \cite{kniu_bupt_icc_2013_beyondturbo,cxia_hkust_tsp_2018_largelist},
the CRC-aided polar codes decoded by SCL with a sufficiently large
list size ($\geq$16) outperform LDPC codes and turbo codes.

\begin{figure}
\includegraphics[width=88mm]{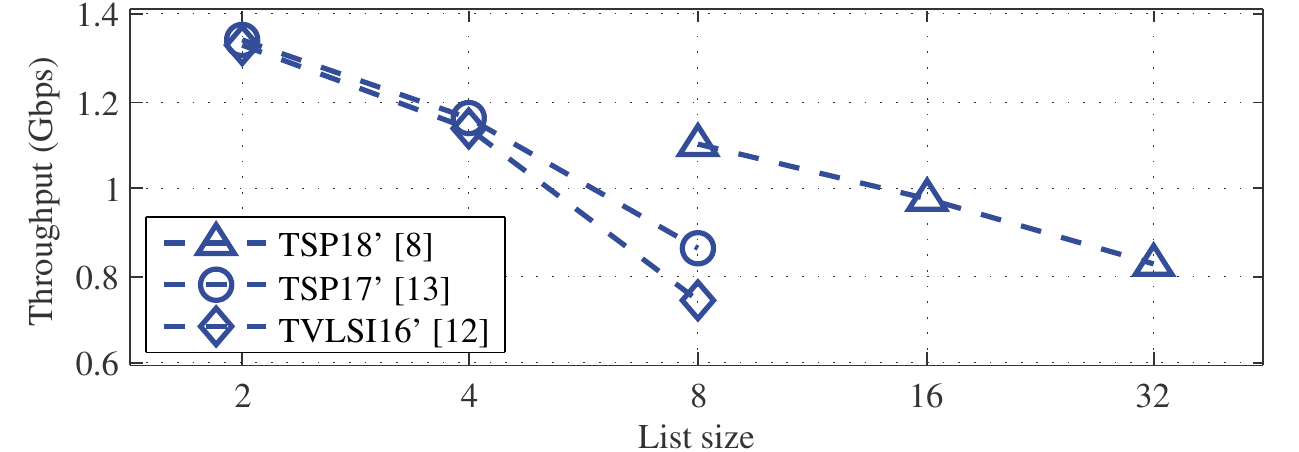}

\caption{Throughputs of various VLSI architectures of SCL decoders synthesized
with or scaled to 90nm CMOS technology. }
\label{fig:soa_thp}
\end{figure}

Due to the extraordinary error correction performance of CRC-aided
SCL decoding, its hardware implementation has attracted much research
interest recently. Several different VLSI architectures \cite{abalatsoukas_epfl_tsp_2015_llrlscd,byuan_umn_tvlsi_2015_sclmbd,crxiong_lehigh_tsp_2016_symbol,jlin_lehigh_tvlsi_2016_highthpt,sahashemi_mcgill_tsp_2017_fastflexible,cxia_hkust_tsp_2018_largelist,pgiard_mcgill_sips_2015_fpga638,crxiong_lehigh_sips_2016_fpgaemul,cxia_hkust_fpl_2017_fpgalarge}
have been proposed for SCL. The decoding throughputs achieved by the
state-of-the-art architectures are shown in \figref \ref{fig:soa_thp}.
It can be observed that the decoding throughputs of all the architectures
are degraded with the list size. This is mainly because the critical
path delay of some of the critical functional modules \cite{abalatsoukas_epfl_iscas_2015_sorting,cxia_hkust_iscas_2018_pathmem,mmousavi_hkust_tsp_2018_lscdps}
in these architectures increases rapidly when the list size is increased.
Although efforts have been made to optimize these modules as well
as the overall architecture, the throughput is still reduced due to
the high decoding complexity. 

To increase the decoding speed so that it can match with that of LDPC
or turbo code architectures, adaptive SCL (A-SCL) decoding was proposed
in \cite{bli_huawei_cl_2012_crc} and a corresponding software decoder
was implemented on CPU in \cite{gsarkis_mcgill_jsac_2016_sclfast}.
This algorithm first uses a single SC to decode a codeword. If the
decoded vector cannot pass CRC, the list size is doubled and the decoding
repeats. This process is iterated until a valid vector is obtained
or a pre-defined $\mathcal{L}_{\text{max}}$ is reached. Experimental
results \cite{bli_huawei_cl_2012_crc} show that A-SCL significantly
reduce the average list size $\bar{\mathcal{L}}$ required to achieve
an equivalent error correction performance of SCL decoding with $\mathcal{L}=\mathcal{L}_{\text{max}}$.
The average throughput of executing A-SCL on hardware can benefit
from the reduction on the $\bar{\mathcal{L}}$. However, if the algorithm
is directly mapped to hardware, the decoding latency of each codeword
is different, which may not support applications that need a constant
input transmission data rate. Also, the hardware complexity is high
as multiple SCL modules are needed.

The main contributions of this work are outlined as follows: 
\begin{enumerate}
\item We simplify the algorithm of A-SCL \cite{bli_huawei_cl_2012_crc}
and propose a two-staged adaptive SCL (TA-SCL) decoding. Different
from \cite{bli_huawei_cl_2012_crc,gsarkis_mcgill_jsac_2016_sclfast},
TA-SCL is more hardware-friendly as it is able to achieve a high throughput
for applications that require a fixed input transmission data rate.
\item An analytical model of TA-SCL is developed based on Markov chain to
analyse its error correction performance. Its accuracy is verified
by simulation, and it can be used for the optimization of the VLSI
architecture for TA-SCL. 
\item Simulation results show that the throughput of TA-SCL with $\mathcal{L}_{\text{max}}=32$
is two times higher than that of the SCL decoder with $\mathcal{L}=32$
\cite{cxia_hkust_tsp_2018_largelist} for good channel conditions
with negligible performance degradation. The throughput is also higher
than those of SCL decoders with smaller list sizes \cite{jlin_lehigh_tvlsi_2016_highthpt,sahashemi_mcgill_tsp_2017_fastflexible}.
\end{enumerate}

\section{Miscellaneous}

\subsection{Introduction of Polar Codes and SCL}

Polar codes are a family of block codes \cite{earikan_bilkent_tit_2009_polar}
characterised by an $N\times N$ binary generator matrix $\mathbf{F}_{N}$,
where $N$ is the code length.%
{} The source word $\textbf{u}_{N}$ and codeword $\textbf{x}_{N}$
of an $N$-bit frame are both binary vectors, and the encoding can
be expressed as $\textbf{x}_{N}=\textbf{u}_{N}\cdot\mathbf{F}_{N}$.
Among all the $N$ bits in a frame, only $K$ bits are used to send
information and the rest are frozen bits which are set to 0. The last
$r$ information bits are used to transmit the checksum of the CRC
code. 
\begin{figure}
\includegraphics[width=88mm]{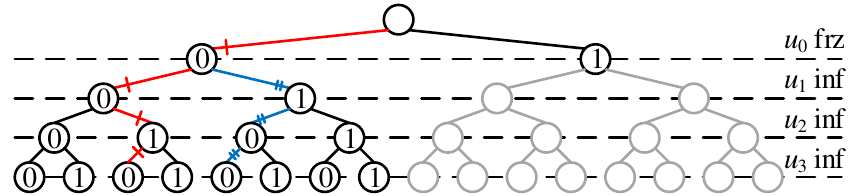}

\caption{The decoding tree of a polar code whose $N=4$ and $\mathcal{L=}2$.}
\label{fig:decoding_tree}
\end{figure}
SCL decoding of polar codes decodes a codeword bit-by-bit in a serial
order, and the decoding process is similar to a search problem on
a binary decoding tree whose depth is $N+1$. A decoding tree with
$N=4$ is shown in \figref \ref{fig:decoding_tree}. The $i^{th}$
source bit $u_{i}$ is mapped to the nodes of the decoding tree at
level $i+1$. A path from the root node to a leaf node represents
a candidate of decoded vector. For a parent at level $i$, its left
and right children at level $i+1$ correspond to the expansions of
the decoding path with $u_{i}=0$ and 1, respectively. In the example,
the path marked with single crosslines represents a decoding vector
$0010$. If a bit, such as $u_{0}$ in \figref \ref{fig:decoding_tree},
is a frozen bit, the sub-tree rooted at the right child does not contain
any valid candidate and hence is pruned. Therefore, the total number
of the possible candidates in a decoding tree is $2^{K}$, and it
is is too large to exhaustively search the decoding tree to obtain
the correct decoded vector when a practical code length is used. To
limit the computational complexity, each SCL decoding has a pre-defined
list size $\mathcal{L}$. %
{} If the number of paths at a certain level exceeds $\mathcal{L}$,
a list management operation is used to select and keep the best $\mathcal{L}$
survival paths and discard the rest ones. The example in \figref
\ref{fig:decoding_tree} maintains a list with $\mathcal{L}=2$, so
another path marked by double crosslines representing the decoded
vector 0100 is also kept in the list. At the end of the decoding,
the path in the list that passes CRC is selected as the output vector.
\vspace*{-12pt}
\begin{small}
\noindent\noindent
	\begin{algorithm}
	\caption{Adaptive SCL with CRC}	
		\textbf{Input:} $N$ channel LLRs; \textbf{Initial:} $\mathcal{L}=1$\;
		\While{$\mathcal{L}\leq\mathcal{L}_{\text{max}}$}{
			SCL with $\mathcal{L}$: codeword from channel\;
			\If{$\geq 1$ paths pass CRC}{Output the most reliable path; Break;}
			\Else{$\mathcal{L}=2\cdot\mathcal{L}$;}
		}	
	\end{algorithm}
\end{small}
\vspace*{-12pt}
\subsection{Adaptive SCL with CRC}
Adaptive SCL with CRC was proposed in \cite{bli_huawei_cl_2012_crc}
and its operation is summarised in Algorithm 1. Each time, a new codeword
which contains $N$ log-likelihood ratios (LLRs) of the input values
is sent for decoding. A-SCL starts from an SCL with $\mathcal{L}=1$,
i.e. a single SC. If there is at least one decoded vectors that pass
CRC at the end of decoding, the one with the highest reliability is
chosen as output. Otherwise, the list size is doubled and the codeword
is decoded again by an SCL with the new list size. Usually, a pre-defined
$\mathcal{L}_{\text{max}}$ is used to limit the computational complexity,
that is, after the decoding using an SCL with $\mathcal{L}_{\text{max}}$,
the decoding terminates even when there is no valid candidate. According
to \cite{bli_huawei_cl_2012_crc}, the error correction performance
of A-SCL is the same as that of an SCL with $\mathcal{L}_{\text{max}}$.
At the same time, as most of the valid decoded vectors can be obtained
using SCL with smaller list sizes, the average list size $\bar{\mathcal{L}}$
of A-SCL is much smaller than $\mathcal{L}_{\text{max}}$ and its
average decoding speed is much higher than that of SCL with $\mathcal{L}_{\text{max}}$.%

\subsection{Problems of Implementing A-SCL on Hardware \label{subsec:problems}}

If the A-SCL algorithm is implemented on hardware, the throughput
will be much higher than that of a traditional SCL. However, direct
mapping of the A-SCL algorithm onto a VLSI architecture requires the
architecture to support multiple SCL decodings with all $\mathcal{L}\in\{1,2,4,...,\frac{\mathcal{L}_{\text{max}}}{2},\mathcal{L}_{\text{max}}\}$.
This increases the design effort and also the hardware complexity.
Moreover, different codewords may need SCL with different list sizes
and SCL with a larger list size has a much higher latency. When a
codeword needs longer latency to decode, the input has to be interrupted
until the decoding of the current frame is finished. Because of that,
a directly-mapped architecture may not be able to support applications
that need to have a constant input data rate, such as the channel
coding blocks in communication networks.

To improve the decoding speed on hardware, a CPU-based software A-SCL
decoder was proposed in \cite{gsarkis_mcgill_jsac_2016_sclfast} in
which A-SCL was simplified by only using a single SC and an SCL with
$\mathcal{L}_{\text{max}}$. However, the variable decoding latency
issue has not yet been addressed. Moreover, the overall latency is
very large as the latency for the data movement between the memory
and the computing resources is dominant. Hence, neither the original
A-SCL nor the simplified A-SCL in \cite{gsarkis_mcgill_jsac_2016_sclfast}
is a good choice for high-throughput VLSI implementations. To solve
these issues and map A-SCL to a high-speed and efficient VLSI architecture,
we propose a two-staged adaptive SCL which will be presented in the
next section.

\section{Two-staged Adaptive SCL\label{sec:ta-scl}}

\subsection{Algorithm of TA-SCL}

\begin{figure}
\includegraphics[width=88mm]{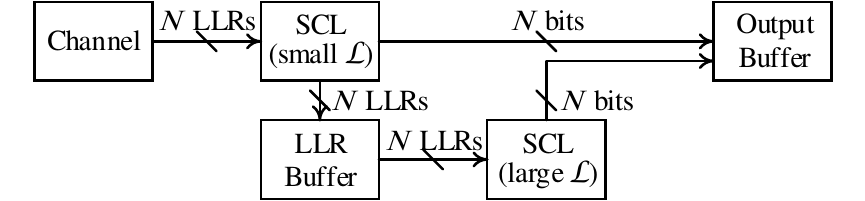}\caption{Block diagram of TA-SCL.}
\label{fig:block_diagram}
\end{figure}
\begin{figure}
\includegraphics[width=88mm]{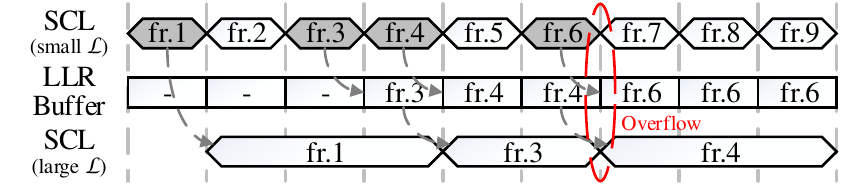}

\caption{Timing schedule of TA-SCL. The codewords in gray cannot be decoded
correctly by $\mathbf{D}_{s}$. The circle shows a buffer overflow.}

\label{fig:schedule}
\end{figure}
As mentioned above, the average list size $\bar{\mathcal{L}}\ll\mathcal{L}_{\text{max}}$
in A-SCL. Actually, $\bar{\mathcal{L}}\approx1$ in a high signal-to-noise
ratio (SNR) operation region \cite{bli_huawei_cl_2012_crc}, indicating
that most codewords are decoded by the single SC correctly. At the
same time, the error correction performance follows that of SCL with
$\mathcal{L}_{\text{max}}$. Based on these observations, we propose
a hardware-friendly two-staged adaptive SCL.%
The block diagram of TA-SCL and its timing schedule are shown in \figref
\ref{fig:block_diagram} and \figref \ref{fig:schedule}, respectively.
Basically, it includes two SCL decodings, which are an SCL decoding
with small list size (not necessarily to be 1), denoted as $\mathbf{D}_{s}$,
and an SCL decoding with large list size, denoted as $\mathbf{D}_{l}$.
Each codeword from the channel is first decoded by $\mathbf{D}_{s}$.
Most of the time, the decoded vector can be decoded correctly. If
none of the candidates in the list passes CRC after this decoding,
e.g. fr.1 in \figref \ref{fig:schedule}, the current codeword will
be decoded again by $\mathbf{D}_{l}$. This decoding usually takes
longer time than decoding using $\mathbf{D}_{s}$. However, different
from A-SCL, $\mathbf{D}_{s}$ will bring in and decode the next codeword
from the channel input immediately instead of waiting for $\mathbf{D}_{l}$
to finish decoding the current codeword. The continuous running of
$\mathbf{D}_{s}$ permits the data to be transmitted at a constant
data rate which is equal to the decoding speed of $\mathbf{D}_{s}$,
while the decoding performance is guaranteed by $\mathbf{D}_{l}$.
Also, the hardware complexity of TA-SCL is effectively reduced as
only two SCL decoders are needed.

If the channel is subject to burst errors, it is possible that a new
codeword cannot be correctly decoded by $\mathbf{D}_{s}$ and the
decoding in $\mathbf{D}_{l}$ has not finished yet. To deal with this,
an LLR buffer is needed to store the LLRs of the codeword from $\mathbf{D}_{s}$
temporarily, such as fr.3 and fr.4 shown in \figref \ref{fig:schedule}.
An output buffer is also needed to re-order the decoded vectors as
the codeword may be decoded out of order. For example, fr.7$\sim$fr.9
are stored in the output buffer until the decoding of fr.6 finishes.

\subsection{Error Correction Performance of TA-SCL}

To analyse the error correction performance of the TA-SCL decoding,
we define its parameters as follows. %
\begin{itemize}
\item $\mathcal{L}_{s}/\mathcal{L}_{l}$: list sizes of $\mathbf{D}_{s}$
/$\mathbf{D}_{l}$ .
\item $\epsilon_{s}/\epsilon_{l}$: BLERs of $\mathbf{D}_{s}$ /$\mathbf{D}_{l}$
.
\item $t_{s}/t_{l}$: decoding time of each codeword using $\mathbf{D}_{s}$
/$\mathbf{D}_{l}$ .
\item $\beta$: speed gain, which is defined as $\frac{t_{l}}{t_{s}}$.
With out loss of generality, we assume $\beta\in\mathbb{Z}^{+}$.
\item $\zeta$: size of the LLR buffer, which equals to the number of codewords
that can be stored in the buffer.
\end{itemize}
We also denote a TA-SCL decoding whose speed gain is $\beta$ and
buffer size is $\zeta$ as $\mathbf{D}_{\text{TA}}(\beta,\zeta)$.
The TA-SCL decoding in the example shown in \figref \ref{fig:schedule}
hence can be described as $\mathbf{D}_{\text{TA}}(3,1)$ and the corresponding
$\mathbf{D}_{l}$ needs $3t_{s}$ to decode a codeword. When a new
codeword needs to be stored in the LLR buffer but the buffer is full
and decoding in $\mathbf{D}_{l}$ has not finished yet, buffer overflow
happens, which will lead to performance degradation for $\mathbf{D}_{\text{TA}}$\footnote{To deal with buffer overflow, either the codeword in $\mathbf{D}_{l}$
or the new one should be thrown away. In the following, we just analyse
the former case and the latter case can be analysed in the same way. }. An example of buffer overflow is marked in \figref \ref{fig:schedule}.
Thus, the BLER of $\mathbf{D}_{\text{TA}}$, denoted as $\epsilon_{\mathbf{D}_{\text{TA}}}$,
is bounded by
\begin{equation}
\epsilon_{l}\leq\epsilon_{\mathbf{D}_{\text{TA}}}<\epsilon_{l}+\text{Pr(Overflow)}.\label{eq:bler_d2}
\end{equation}
Obviously, it is important to prevent the buffer overflow in order
to reduce $\epsilon_{\mathbf{D}_{\text{TA}}}$. A large buffer size
$\zeta$ certainly helps as more codewords can be stored, and a smaller
speed gain $\text{\ensuremath{\beta}}$ indicates $\mathbf{D}_{l}$
have relatively more time to decode the codewords accumulated in the
buffer. To obtain the best tradeoff among performance, hardware usage
and throughput, an analytical model of $\mathbf{D}_{\text{TA}}$ will
be introduced to derive the relationship between $\text{Pr(Overflow)}$
and the parameters of $\mathbf{D}_{\text{TA}}$ in the next sub-section.%

\subsection{Analytical Model of TA-SCL based on Markov Chain}

\begin{figure}
\subfloat[]{\includegraphics[scale=1]{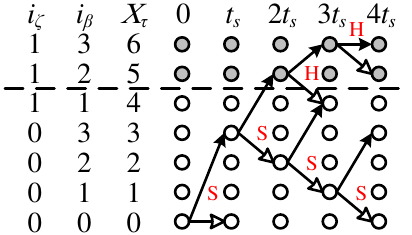}}\subfloat[]{\includegraphics[scale=0.95]{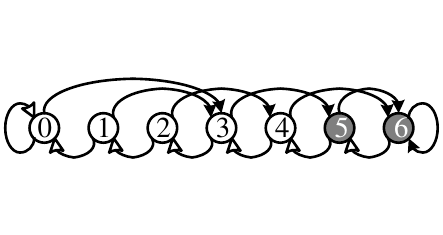}

}

\caption{(a) States and state transitions of $\mathbf{D}_{\text{TA}}(3,1)$
and (b) the corresponding state diagram. The white and black arrows
mean the frame is decoded correctly and incorrectly, respectively.}

\label{fig:state_diagram}
\end{figure}
To model the behavior of $\mathbf{D}_{\text{TA}}(\beta,\zeta)$, we
first introduce the states that the decoder can operate at. In particular,
these states reflect whether buffer overflow will happen. We define
the number of codewords stored in the LLR buffer as $i_{\zeta}$ and
the remaining time required to finish the decoding of $\mathbf{D}_{l}$
(in term of $t_{s}$) as $i_{\beta}$. Each codeword in the LLR buffer
needs $\beta t_{s}$ to decode. Then, the state of TA-SCL indicates
the time to clear the buffer and is defined as
\begin{equation}
X_{\tau}=\beta\cdot i_{\zeta}+i_{\beta},\,i_{\zeta}\in[0,\zeta],\,i_{\beta}\in[0,\beta],
\end{equation}
which equals to the total time required to clear the buffer. For a
$\mathbf{D}_{\text{TA}}(\beta,\zeta)$, there are totally $\mathcal{S}=\beta\zeta+\beta+1$
states. All the $\mathcal{S}$ states can be divided into two groups.
\begin{itemize}
\item Hazard states: The states that the LLR buffer is full and the current
codeword decoded by $\mathbf{D}_{l}$ cannot be finished within $t_{s}$,
which means $i_{\zeta}=\zeta$ and $i_{\beta}>1$. Buffer overflow
will occur if $\mathbf{D}_{s}$ cannot decode the next codeword correctly.
\item Safe states: In contrast with the hazard states, these states do not
have overflow hazard as the LLR buffer has enough space for a codeword
that cannot be correctly decoded by $\mathbf{D}_{s}$.
\end{itemize}
We show an example for $\mathbf{D}_{\text{TA}}(3,1)$ in \figref
\ref{fig:state_diagram}(a), where the black and white arrows represent
the probabilities of $\epsilon_{s}$ and $\epsilon_{s}'=1-\epsilon_{s}$,
respectively. The first three columns show $i_{\beta}$, $i_{\zeta}$
and $X_{\tau}$, respectively. Typical transitions from hazard and
safe states are marked with ``H'' and ``S'' in the figure, respectively.
Note that the transition from state 0 is a little different as $\mathbf{D}_{l}$
is idle. %

Suppose that $\epsilon_{s}$ (BLER of $\mathbf{D}_{s}$) follows an
identical and independent distribution (IID). Then, the state transitions
only depend on current state of $\mathbf{D}_{\text{TA}}(\beta,\zeta)$
and $\epsilon_{s}$. Hence, decoding with $\mathbf{D}_{\text{TA}}$
is a Markov process and can be modeled with a Markov chain. The state
diagram of a $\mathbf{D}_{\text{TA}}(\beta,\zeta)$ can be easily
obtained by finding out all the possible state transitions in \figref
\ref{fig:state_diagram}(a). \figref \ref{fig:state_diagram}(b)
shows the state diagram of $\mathbf{D}_{\text{TA}}(3,1)$. For further
mathematical analysis, we map the state diagram to a transition matrix
$P$ whose size is $\mathcal{S}\times\mathcal{S}$. An element $P_{x.y}\in P$
$(x.y\in[0,\mathcal{S}-1])$ corresponds to the transition probability
from state $x$ to state $y$, i.e.,
\begin{equation}
P_{x.y}=\text{Pr}(X_{\tau+1}=y|X_{\tau}=x),
\end{equation}
where $X_{\tau}$ is the current state and $X_{\tau+1}$ is the next
state. The transition matrix of $\mathbf{D}_{\text{TA}}(3,1)$ mapped
from the state diagram is{\small{}
\begin{equation}
\begin{bmatrix}\epsilon_{s}' & 0 & 0 & \epsilon_{s} & 0 & 0 & 0\\
\epsilon_{s}' & 0 & 0 & \epsilon_{s} & 0 & 0 & 0\\
0 & \epsilon_{s}' & 0 & 0 & \epsilon_{s} & 0 & 0\\
0 & 0 & \epsilon_{s}' & 0 & 0 & \epsilon_{s} & 0\\
0 & 0 & 0 & \epsilon_{s}' & 0 & 0 & \epsilon_{s}\\
0 & 0 & 0 & 0 & \epsilon_{s}' & 0 & \epsilon_{s}\\
0 & 0 & 0 & 0 & 0 & \epsilon_{s}' & \epsilon_{s}
\end{bmatrix}.
\end{equation}
}{\small \par}

With the transition matrix $P$, we can do steady-state analysis for
$\mathbf{D}_{\text{TA}}$. Suppose that the decoding begins with $\mathbf{D}_{\text{TA}}$
at state 0, i.e., the state probability $\lambda_{0}=[1,0,...,0]$.
After $k\cdot t_{s}$ ($k\in\mathbb{Z}^{+}$), the state probability
becomes $\lambda_{k}=\lambda_{0}\cdot P^{(k)}$. Define $P_{\infty}=\lim_{k\rightarrow\infty}P^{(k)}$,
then the steady-state distribution $\lambda_{\infty}$ of $\mathbf{D}_{\text{TA}}$
is
\begin{equation}
\lambda_{\infty}=\lambda\cdot P_{\infty}=\{(P_{\infty})_{0,0},...,(P_{\infty})_{0,\beta\zeta+\beta}\}.\label{eq:ss_dist}
\end{equation}
Actually, all the lines of $P_{\infty}$ are the same, which means
the steady-state distribution is irrespective of the initial state
$\lambda_{0}$ of $\mathbf{D}_{\text{TA}}$. Buffer overflow happens
when $\mathbf{D}_{\text{TA}}$ is in any hazard state and $\mathbf{D}_{s}$
cannot decode the next codeword correctly, and the probability of
buffer overflow is then expressed as
\begin{eqnarray}
\text{Pr(Overflow)} & = & \epsilon_{s}\cdot\text{Pr}(i_{\zeta}=\zeta\text{ and }i_{\beta}>1)\\
 & = & \epsilon_{s}\cdot\text{Pr}(X_{\tau}>\beta\zeta+1),\\
 & = & \epsilon_{s}\cdot{\displaystyle \sum_{i=\beta\zeta+2}^{\beta\zeta+\beta}}(\lambda_{\infty})_{i}.\label{eq:overflow2}
\end{eqnarray}
This probability of overflow bounds $\epsilon_{\mathbf{D}_{\text{TA}}}$
in \eqref{eq:bler_d2}. It is a function of error correction performance
$\epsilon_{s}$, speed gain $\beta$ and buffer size $\zeta$, i.e.,
$\text{Pr(Overflow)}$=$f(\epsilon_{s},\beta,\zeta)$. If $\beta$
and $\zeta$ are fixed, the $\Sigma$ term and hence $\text{Pr(Overflow)}$
is monotonically increasing with respect to $\epsilon_{s}$. The proof
is omitted due to page limitation and will be given in our future
work. The monotonicity indicates we can either increase $\mathcal{L}_{s}$
or the SNR to get a better error correction performance.

We will show the accuracy of the proposed model by simulation results
in the next section. We will also show that TA-SCL can improve the
decoding throughput with a small hardware overhead.

\section{Experimental Results}

\subsection{Accuracy of the Proposed Model}

\begin{figure}
\includegraphics[width=88mm]{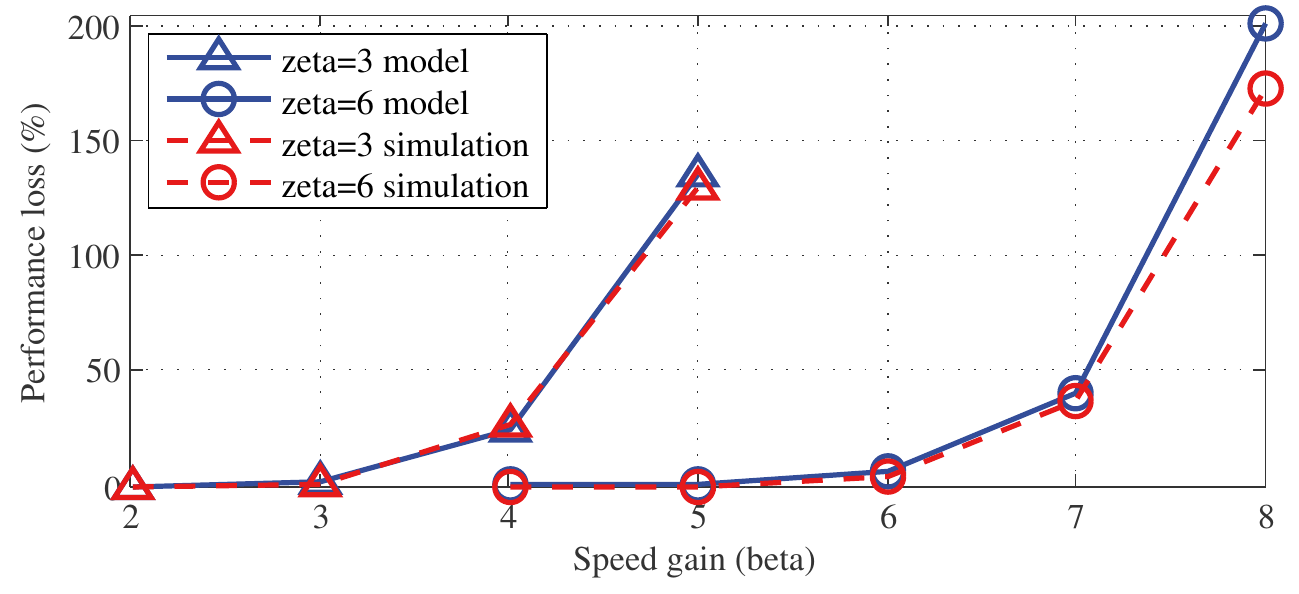}

\caption{Performance loss calculated by model and from simulation.}
\label{fig:perf_loss}
\end{figure}
To verify the accuracy of the proposed model, we run simulations for
a polar code with $\{N,K,r\}=\{1024,512,24\}$ under AWGN channel
conditions. The list sizes of the two component SCL decoders are $\mathcal{L}_{s}=1$
and $\mathcal{L}_{l}=32$, respectively. The simulated BLER results
of $\mathbf{D}_{\text{TA}}$ under different speed gain and buffer
sizes are obtained at an SNR of 2dB and are compared with the upper
bounds calculated using \eqref{eq:bler_d2} and \eqref{eq:overflow2}. 

\figref \ref{fig:perf_loss} summarizes the performance loss with
respect to the speed gain $\beta$ when different $\zeta$ are used.
Here, the performance loss is calculated by $\frac{\epsilon_{\textbf{D}_{\text{TA}}}-\epsilon_{l}}{\epsilon_{l}}\cdot100\%$.
The solid lines and the dashed lines show the calculated and simulated
results, respectively. It can be seen that these two lines are almost
overlapped, indicating $\epsilon_{\mathbf{D}_{\text{TA}}}$ is approximately
equal to its upper bound derived in \eqref{eq:overflow2}. The proposed
model can thus be used to estimate the error correction performance
of an $\mathbf{D}_{\text{TA}}$ accurately. The results also show
that a larger buffer size enables the decoding to run at a higher
speed gain with the same constraint of performance loss. %

\subsection{Analysis of Hardware Gain}

\begin{figure}
\includegraphics[width=88mm]{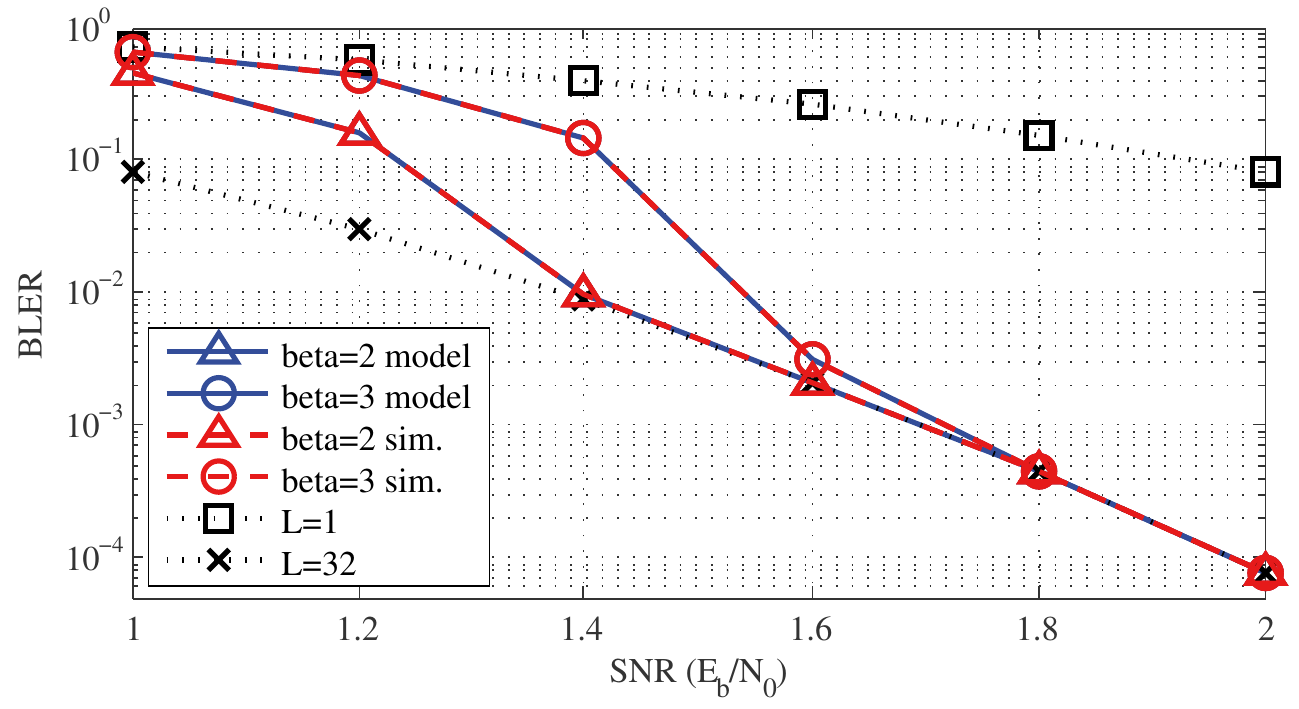}

\caption{Error correction performance of $\mathbf{D}_{\text{TA}}$ with $\zeta=6$.}
\label{fig:err_d2}
\end{figure}
\begin{table}
\caption{Hardware performance of SCL architectures}
\label{tab:soa_arch}

\begin{tabular}{c||c|c|c|c}
\hline 
\multirow{2}{*}{} & \multirow{2}{*}{{\small{}$\mathcal{L}$}} & {\small{}Throughput} & {\small{}Area} & {\small{}Area eff.}\tabularnewline
 &  & {\small{}(Mbps)} & {\small{}($\text{mm}{}^{2}$)} & {\small{}(Mbps/$\text{mm}{}^{2}$)}\tabularnewline
\hline 
{\small{}$\mathbf{D}_{s}$, \cite{pgiard_mcgill_jsps_2018_lowrate}} & {\small{}1} & {\small{}2686$\star$} & {\small{}1.32$\star$} & {\small{}2035}\tabularnewline
\hline 
{\small{}$\mathbf{D}_{l}$, \cite{cxia_hkust_tsp_2018_largelist}} & {\small{}32} & {\small{}827} & {\small{}19.58} & {\small{}42}\tabularnewline
\hline 
{\small{}Proposed} & \multirow{2}{*}{{\small{}32}} & {\small{}2481 ($\geq$1.6dB)} & \multirow{2}{*}{{\small{}23.12}} & {\small{}110}\tabularnewline
\cline{3-3} \cline{5-5} 
{\small{} $\mathbf{D}_{\text{TA}}$ (Est.)} &  & {\small{}1654 ($\geq$1.4dB)} &  & {\small{}73}\tabularnewline
\hline 
\multicolumn{5}{l}{{\small{}$\star$Scaled to 90 nm technology.}}\tabularnewline
\end{tabular}
\end{table}
 In this sub-section, we show the improvement of hardware performance
achieved by the proposed TA-SCL decoder. We use a polar code with
$\{N,K,r\}$=$\{1024,512,24\}$ and $\mathcal{L}$=32 as an example.
The hardware performance of some VLSI architectures of SCL decoder
in the literature \cite{cxia_hkust_tsp_2018_largelist,pgiard_mcgill_jsps_2018_lowrate}
is shown in \tabref \ref{tab:soa_arch}. They are used as the component
SCL decoders in the TA-SCL decoder. \figref \ref{fig:err_d2} shows
the error correction performance of $\mathbf{D}_{\text{TA}}$ with
buffer size $\zeta$=6. When the target $\beta$ is 3, there is almost
no performance degradation at a high SNR range ($\geq$1.6dB) comparing
with the baseline of $\mathcal{L}$=32. The degradation is obvious
at a low SNR range. If the target $\beta$ is reduced to 2, the decoder
can work in a wider range of SNR down to 1.4dB. All these observations
is consistent with the intuitions mentioned in \secref IV. It is
noted that the throughput of $\mathbf{D}_{\text{TA}}$ is lower than
$\mathbf{D}_{s}$ in both cases, so the speed gain $\beta$ of up
to 3x is achievable. The overall throughput of TA-SCL is also higher
than that of the SCL decoders with smaller list sizes as shown in
\figref \ref{fig:soa_thp} \cite{jlin_lehigh_tvlsi_2016_highthpt,sahashemi_mcgill_tsp_2017_fastflexible}.

The area of the proposed architecture is shown in \tabref \ref{tab:soa_arch},
which is estimated based on the results reported in the literature
\cite{cxia_hkust_tsp_2018_largelist,pgiard_mcgill_jsps_2018_lowrate}.
It equals to the sum of area of the two SCL modules, the LLR buffer
and the output buffer. As the area of the $\mathbf{D}_{l}$ module
is dominant, the proposed $\mathbf{D}_{\text{TA}}$ only has a 18\%
area overhead. Moreover, due to the throughput improvement, the area
efficiency of $\mathbf{D}_{\text{TA}}$ is also much higher than that
of $\mathbf{D}_{l}$.

\section{Conclusion}

In this work, a two-staged adaptive SCL is proposed. This algorithm
can support data input at a fixed data rate and has a low hardware
complexity. To analyse its error correction performance, an analytical
model is also proposed and its accuracy is then verified by simulations.
With a good selection of the parameters of TA-SCL using the proposed
analytical model, an optimal tradeoff between speed gain, error correction
performance loss and hardware overhead can be obtained for designing
the VLSI architecture.

\end{document}